\documentclass[aps,prl,reprint,superscriptaddress,amsmath,amssymb]{revtex4-2}
\usepackage{xcolor}
\definecolor{prlblue}{RGB}{29, 29, 129}
\usepackage[colorlinks=true, citecolor=prlblue, linkcolor=prlblue, urlcolor=prlblue]{hyperref}
\usepackage{graphicx}
\usepackage{dcolumn}
\usepackage{bm}
\usepackage{color}
\usepackage{comment}

\begin{document}


\title{Critical angles and one-dimensional moir\'e physics in twisted rectangular lattices}

\author{Dongdong An}
\thanks{These authors contributed equally to this work.}
\affiliation{National Laboratory of Solid-State Microstructures, Collaborative Innovation Center of Advanced Microstructures, School of Physics, Nanjing University, Nanjing, China}

\author{Tao Zhang}
\thanks{These authors contributed equally to this work.}
\affiliation{Department of Materials Science and Engineering, City University of Hong Kong, 99977 Kowloon, Hong Kong}
\affiliation{Songshan Lake Materials Laboratory, 523808 Dongguan, Guangdong, China} 
\affiliation{Tsientang Institute for Advanced Study, Zhejiang 310024, China}

\author{Qiaoling Xu}
\affiliation{College of Physics and Electronic Engineering, Center for Computational Sciences, Sichuan Normal University, Chengdu 610068, China}
\affiliation{Tsientang Institute for Advanced Study, Zhejiang 310024, China} 

\author{Hailing Guo}
\affiliation{Tsientang Institute for Advanced Study, Zhejiang 310024, China} 
\affiliation{Department of Engineering, University of Cambridge, Cambridge CB2 1PZ, United Kingdom}

\author{\\Majeed Ur Rehman}
\affiliation{Songshan Lake Materials Laboratory, 523808 Dongguan, Guangdong, China}
\affiliation{Tsientang Institute for Advanced Study, Zhejiang 310024, China} 

\author{Dante M. Kennes}
\affiliation{Institut f\"ur Theorie der Statistischen Physik, RWTH Aachen University and JARA-Fundamentals of Future Information Technology, 52056 Aachen, Germany}
\affiliation{Max Planck Institute for the Structure and Dynamics of Matter, Luruper Chaussee 149, 22761 Hamburg, Germany} 


\author{Angel Rubio}
\affiliation{Max Planck Institute for the Structure and Dynamics of Matter, Luruper Chaussee 149, 22761 Hamburg, Germany} 

\author{Lei Wang}
\email[Email: ]{leiwang@nju.edu.cn}
\affiliation{National Laboratory of Solid-State Microstructures, Collaborative Innovation Center of Advanced Microstructures, School of Physics, Nanjing University, Nanjing, China}

\author{Lede Xian}
\email[Email: ]{ldxian@tias.ac.cn}
\affiliation{Tsientang Institute for Advanced Study, Zhejiang 310024, China} 
\affiliation{Songshan Lake Materials Laboratory, 523808 Dongguan, Guangdong, China} 
\affiliation{Max Planck Institute for the Structure and Dynamics of Matter, Luruper Chaussee 149, 22761 Hamburg, Germany} 

\date{\today}

\begin{abstract}
        Engineering moir\'e superlattices in van der Waals heterostructures provides fundamental control over emergent electronic, structural, and optical properties allowing to affect topological and correlated phenomena. This control is achieved  through imposed periodic modulation of potentials and targeted modifications of symmetries. 
        For twisted bilayers of van der Waals materials with rectangular lattices, such as PdSe$_2$, this work shows that one-dimensional (1D) moir\'e patterns emerge universally. This emergence is driven by a series of critical twist angles (CAs). We investigate the geometric origins of these unique 1D moir\'e patterns and develop a universal mathematical framework to predict the CAs in twisted rectangular lattices. Through a density functional theory (DFT) description of the electronic properties of twisted bilayer PdSe$_2$, we further reveal directionally localized flat band structures, localized charge densities and strong spin-orbit coupling along the dispersive direction which points to the emergence of an effectively 1D strongly spin-orbit coupled electronic systems. This establishes twisted rectangular systems as a unique platform for engineering low-symmetry moir\'e patterns, low-dimensional strongly correlated and topological physics, and spatially selective quantum phases beyond the isotropic paradigms of hexagonal moir\'e materials.

\end{abstract}

\maketitle
\textit{Introduction}---moir\'e superlattices emerge in vertically stacked two-dimensional (2D) material heterostructures when adjacent layers are rotated relative to each other by an angle $\theta$. This rotation usually produces a long-ranged periodic modulation in both in-plane directions that far exceeds the atomic lattice scale \cite{oden1991superperiodic,PRB1993Moire,PRL2007TBGEleStru,suarez2010flat,MacDonald2011}. Such moir\'e-induced modulations dramatically reshape the electronic structure, giving rise to flat bands and, comparably speaking, enhanced electron-electron interaction effects \cite{PRL2007TBGEleStru,MacDonald2011,PRM_hu2017DiracEle,wang2025moire,PRB2022Topo&Flat,NP2020SC&Co}. These features have enabled the discovery of a wide range of correlated and topological phases, including unconventional superconductivity, correlated insulators, and fractional quantum anomalous Hall states \cite{NP_Tong2017Topo,Sci_spanton2018ObserFracChern,Cao2018unconven,Wang2020Corre,PRB_kundu2022MoireFlat,PRL_ghorashi2023Topo&Flat_BG,PRL2024QAHC,PRR_2025TopoExcitons}. However, still today the rich interplay between lattice geometry, electronic bandwidth, and Coulomb interactions in 2D moir\'e systems continues to drive fundamental advances in understanding emergent quantum phenomena in van der Waals materials \cite{PRR_carr2020duality&Bloch,NM_andrei2020TBG,N_regan2020Mott&WCS,NP_kennes2021MoireInCMQ,PRB_tang2021GeomOriofTopoIn,arX_xu2025UnSCNearFQAHI,NC_jiang2025GateTunFlat}.

While early research on moir\'e materials mainly focused on twisted 2D materials with triangular or hexagonal lattices, such as twisted bilayer or multilayer graphene and twisted transition metal dichalcogenides (TMDs) \cite{Cao2018unconven,Wang2020Corre,Na_liu2020TunableSpinPCorSInTBG,NC_mmiao2021strongInterction,NRP_mak2019ProbingContro,NC_liao2020PreciseContrMoS2,NC_van2023RotAdilationalReconstructionInTMDs}, recent attention has also extended to heterostructures with rectangular lattices \cite{NC_Kennes_2020_1DFlatInGeSe,NC_zhao2021AnisotropicOptInTBP,Na_Wang_2022_1DLLs,PRB_magorrian2024Strain1DInWTe2,PRB_soltero2022moireTBP,PNAS_liu2025MoireCrSBr,Xu_2024Square,SciAd_Hongli2024_1DMoireExcitons}. Unlike triangular systems, which exhibit high-symmetry moir\'e patterns (e.g., $C_3$,$C_6$) due to their threefold or sixfold rotational symmetry, rectangular lattices possess lower rotational and mirror symmetries. Consequently, their moir\'e superlattices typically display reduced symmetries (e.g., $C_1$, $C_2$) as well. This symmetry reduction, however, enables a greater diversity of anisotropic moir\'e patterns and emergent electronic behaviors, offering new avenues for tunable quantum phenomena; a realm of rich physics that remains largely unexplored. 

\begin{figure*}
\includegraphics[width=0.9\linewidth,trim=0 0 0 0,clip]{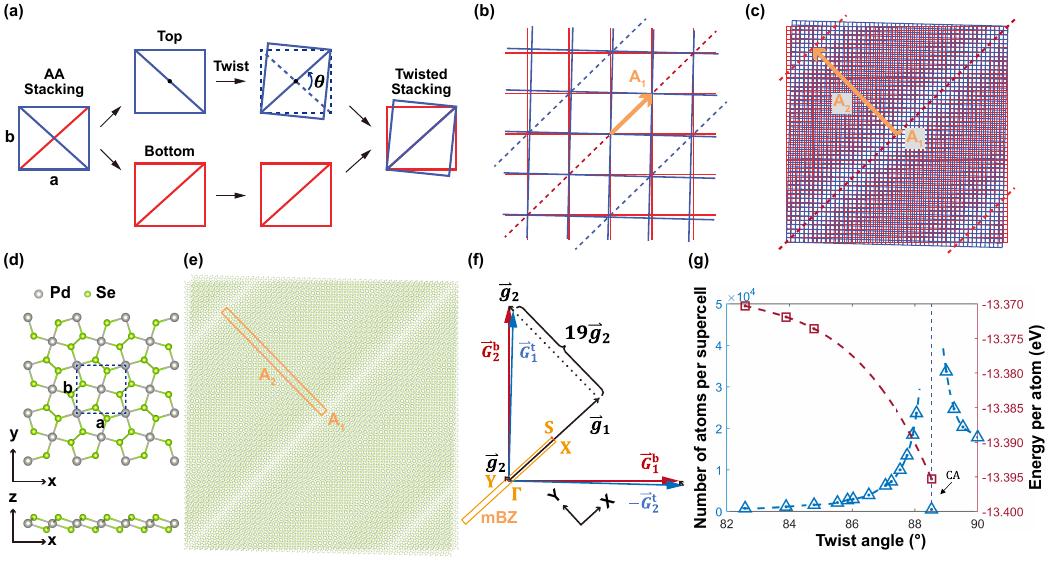}
\caption{
Emergence of 1D moir\'e patterns at critical twist angles in rectangular lattices. 
(a) Schematic of twisted bilayer rectangular lattice at the first critical angle $\theta_{\text{c1}}$.  
(b,c) Emerging 1D moir\'e superlattice showing the atomic (b) and the long-range moir\'e periodicity (c). The two supercell lattice vectors, $\bm{A}_{1}$ and $\bm{A}_{2}$, are highlighted in orange. (d) Top and side views of the atomic structure of monolayer PdSe$_2$. The dashed rectangle denotes the primitive cell. (e) Formation of a 1D moir\'e pattern in twisted bilayer PdSe$_2$ at $\theta_{\text{c1}}= 88.53^\circ$. The solid orange rectangle marks the moir\'e supercell. (f) Reciprocal space of twisted bilayer PdSe$_2$. $\bm{G}_{1}^\text{t}$, $\bm{G}_{2}^\text{t}$ (top layer) and  $\bm{G}_{1}^\text{b}$, $\bm{G}_{2}^\text{b}$ (bottom layer) are primitive reciprocal vectors.  The moir\'e reciprocal vectors $\bm{g}_\text{1}$, $\bm{g}_\text{2}$ are defined by $\bm{g}_\text{1}$ = $\bm{G}_{1}^\text{t}-19\bm{g}_\text{2}$ and $\bm{g}_\text{2}$ = $\bm{G}_{2}^\text{b}-\bm{G}_{1}^\text{t}$. The orange rectangle indicates the moir\'e Brillouin zone (mBZ). Total energy from DFT calculations (red) and the minimal number of atoms per supercell (blue) as a function of twist angle near $\theta_{\text{c1}}=88.53^\circ$. Dashed lines serve as guides to the eye.}
\label{fig:figure1}
\end{figure*}

In twisted triangular and hexagonal moir\'e systems, the moir\'e periodicity (and thus the number of atoms per moir\'e cell) scales inversely with the twist angle \cite{PRL2007TBGEleStru,feuerbacher2021moire}. While this behavior is often assumed to be universal, in this work, we find that for twisted 2D rectangular lattices, there exists a series of critical twist angles at which the number of atoms in the moir\'e supercell dramatically drops. Using twisted bilayer PdSe$_2$ as an example, we show that these specific twist angles correspond to the formation of 1D moir\'e stripes in the system \cite{Sinner2023Strain1D}. We elucidate the structural origins of these anisotropic moir\'e patterns at the critical twist angles and investigate the corresponding electronic structures. We show that at these specific twist angles, the low-energy bands exhibit pronounced quasi-1D characteristics, strong spin-orbit coupling and relatively flat dispersion, establishing an ideal platform for investigating quasi-1D strongly correlated physics, topologically protected edge states and materials with interesting optical excitations \cite{SciAd_Hongli2024_1DMoireExcitons,Na_Wang_2022_1DLLs,klein2023bulk,yu2023evidence,Khoury2024Toward1D}. Our approach can be readily extended to other low-symmetry twisted lattices, providing a theoretical framework for understanding the twist-angle dependence of moir\'e patterns and related phenomena in twisted rectangular and low-symmetry lattices.

\textit{Geometric origin of 1D moir\'e pattern}---For twisted triangular or hexagonal lattices, twisting the top and bottom layers relative to each other by a finite angle that is not a multiple of $\pi/3$ leads to the formation of a two-dimensional periodic moir\'e pattern \cite{PRL2007TBGEleStru,MacDonald2011}. This arises from the disruption of in-plane atomic alignment caused by the twist. Despite this misalignment at the atomic scale, certain lattice sites periodically realign at a larger moir\'e length scale, giving rise to the long-ranged periodicity characteristic of moir\'e patterns. For rectangular lattices with lower symmetry, a distinct and intriguing scenario arises. Due to the equal size of the two diagonals of a rectangle, it is possible to rotate the top layer by a specific angle $\theta$ such that atomic sites realign along one diagonal direction while remaining misaligned in the perpendicular directions (see Fig.~\ref{fig:figure1}(a) for an illustration). This special twisted geometry preserves atomic periodicity along one diagonal direction of the original unit cell, as shown in Fig.~\ref{fig:figure1}(b), while introducing a long-ranged periodic modulation in the orthogonal in-plane directions, as illustrated in Fig.~\ref{fig:figure1}(c). The resulting structure forms an array of one-dimensional stripes -- a one-dimensional moir\'e pattern in a two-dimensional plane. It is important to note that the reduced symmetry arising from unequal lattice constants (${b} \neq {a}$) is essential for the emergence of this behavior. If the lattice were symmetric (${b} = {a}$), such a rotation would preserve atomic periodicity in both in-plane directions, and no moir\'e pattern would form. In the following, we use twisted bilayer PdSe$_2$ as a representative example to illustrate this special configuration and its impact on the electronic properties.
PdSe$_2$ is an air-stable 2D material with a rectangular lattice structure and anisotropic lattice constants that can be synthesized down to the monolayer \cite{oyedele2017pdse2,sun2015electronic} (see Fig.~\ref{fig:figure1}(d)). 
For monolayer PdSe$_2$, the Bravais lattice sites can be described by the lattice vectors:
$    \bm{L}=n_1 \bm{a} +n_2 \bm{b},$ 
where $\bm{a}$ and $\bm{b}$ are the primitive lattice vectors of monolayer PdSe$_2$, and $n_1$, $n_2$ are integers. The twisted bilayer structure is constructed by rotating one layer with respect to the other. We generate the lattice points of the rotated layer by applying a rotation matrix R($\theta$) to the original Bravais lattice vectors. The rotation matrix can be expressed as 
 $   R(\theta) = 
    \begin{pmatrix}
    \cos\theta & -\sin\theta \\
    \sin\theta &  \cos\theta 
    \end{pmatrix} $
where $\theta$ is the twist angle. The Bravais lattice points of the twisted layer can be obtained by
$    \bm{L'} = R(\theta)L =n_1 \bm{a'} +n_2 \bm{b'}, $
 where $\bm{a'}$ and $\bm{b'}$ are the primitive lattice vectors of the twisted layer, and are related to the original vectors $\bm{a}$ and $\bm{b}$ as 
 $   \bm{a'} = R(\theta)\bm{a}, \bm{b'} = R(\theta)\bm{b}.$  
For a lattice point ($p\bm{a'}, q\bm{b'}$) of the rotated layer to align with a lattice point ($m\bm{a}, n\bm{b}$) of the fixed layer after a rotation by angle $\theta$, then there must exist a set of integers ($m, n, p, q$) satisfying the following equation:
\begin{equation}
    p \bm{a'} + q \bm{b'} =  m \bm{a} + n \bm{b}
    \label{eq:tosolve}
\end{equation}
Due to the orthogonality of both $\bm{a}$ and $\bm{b}$ as well as $\bm{a'}$ and $\bm{b'}$, Eq. \eqref{eq:tosolve} yields:
$    (p^2-m^2)a^2 = (n^2-q^2)b^2   $
This equation can be satisfied by
\begin{equation}
    p = \pm m \quad \text{and} \quad q = \pm n 
    \label{eq:sol1}
\end{equation}
or
\begin{equation}
    \sqrt{\frac{p^2-m^2}{n^2-q^2}} = \frac{b}{a}, \quad ( m \neq \pm p \text{ and } n \neq \pm q)
    \label{eq:sol2}
\end{equation}

Although both Eqs.~\eqref{eq:sol1} and \eqref{eq:sol2} can be used to identify lattice points in the top layer that can realign with those in the bottom layer after rotation, Eq.~\eqref{eq:sol1} offers a simpler condition, as $n$ and $m$ can be arbitrary integers, as long as $p$ and $q$ satisfy Eq.~\eqref{eq:sol1}. Excluding trivial solutions, this implies that the lattice vector $\bm{A_1}=m \bm{a} +n \bm{b}=(ma,nb)$ can serve as one of the moir\'e supercell lattice vectors for any nonzero combinations of $n$ and $m$. The possible lattice vectors correspond to the diagonals of an $m\times n$ supercell. The associated twist angle is determined by twice the angle between the diagonal vector and the coordinate axes, given by $\theta_1=2\arctan\left({ma}/{nb}\right)$ or $\theta_2=2\arctan\left({nb}/{ma}\right)$. These two twist angles correspond to the same moir\'e supercell and satisfy $\theta_1+\theta_2=180^\circ$. For simplicity, we focus on the angle closest to $0^\circ$ in this work. For small value of $n$ and $m$, the supercell period along the corresponding lattice vector can be comparable to that of the original primitive cell. The simplest example is given by $\left(m, n, p, q\right) = \left(1, 1, 1, -1\right)$, which yields a supercell lattice vector $\bm{A_1} = \bm{a} + \bm{b} = (a, b)$, corresponding precisely to the configuration illustrated in Fig.~\ref{fig:figure1}(a-c). The corresponding moir\'e superlattice for twisted bilayer PdSe$_2$ at this configuration is shown in Fig.~\ref{fig:figure1}(e), with a twist angle $\theta_{\text{c1}} = 2\arctan\left({b}/{a}\right)=88.53^{\circ}$.

The discussion above provides the mathematical framework for identifying the moir\'e supercell lattice vector $\bm{A_1}$ that retains atomic periodicity. The second lattice vector $\bm{A_2}$, associated with the long-range moir\'e periodicity, can be either identified numerically (see section I in the Supplementary Material) or inferred from the corresponding supercell reciprocal lattice vectors. Fig.~\ref{fig:figure1}(f) presents the reciprocal space of twisted bilayer PdSe$_2$ at $88.53^{\circ}$. In general, the reciprocal lattice vectors of the moir\'e supercell can be constructed from the smallest differences between the primitive reciprocal lattice vectors of the top and bottom layers, i.e., $\bm{g_i}=\bm{G_i}^{b}-\bm{G_j}^{t}$, where $\bm{G_i}^{b}$ and $\bm{G_i}^{t}$ are the primitive cell reciprocal lattice vectors of the bottom and top layers, respectively. However, in this case, it is interesting to note that the vectors $\bm{G_2}^{b}-\bm{G_1}^{t}$ and $\bm{G_1}^{b}-\bm{G_2}^{t}$ are collinear and identical, defining only one of the moir\'e supercell lattice vectors $\bm{g_2}$. Such behavior is similar to that observed in twisted triangular lattices subjected to critical heterostrain, which leads to the formation of ideal one-dimensional moir\'e structures \cite{Sinner2023Strain1D}. The other supercell reciprocal lattice vector $\bm{g_1}$ can be identified as $\bm{g_1}=\bm{G_1}^{t} - 19\bm{g_2}=20\bm{G_1}^{t}-19\bm{G_2}^{b}$, as shown in Fig.~\ref{fig:figure1}(f). The corresponding moir\'e supercell lattice vector $\bm{A_2}$ is then obtained from the relation 
$\bm{A_2}=2\pi\frac{\bm{g_1} \times \hat{\bm{z}}}{\bm{g_1} \cdot (\bm{g_2} \times \hat{\bm{z}})}.$ 
Similarly, it can be confirmed that $\bm{A_1}$ satisfies 
 $   \bm{A_1}=2\pi\frac{\bm{g_2} \times \hat{\bm{z}}}{\bm{g_1} \cdot (\bm{g_2} \times \hat{\bm{z}})},$ 
ensuring that the moir\'e supercell lattice vectors and reciprocal lattice vectors satisfy the orthonormal condition $\bm{A_i}\cdot \bm{g_j}=2\pi\delta_{ij}, (i,j=1,2)$.    

\begin{figure*}
\includegraphics[width=0.9\linewidth,trim=0 0 0 0,clip]{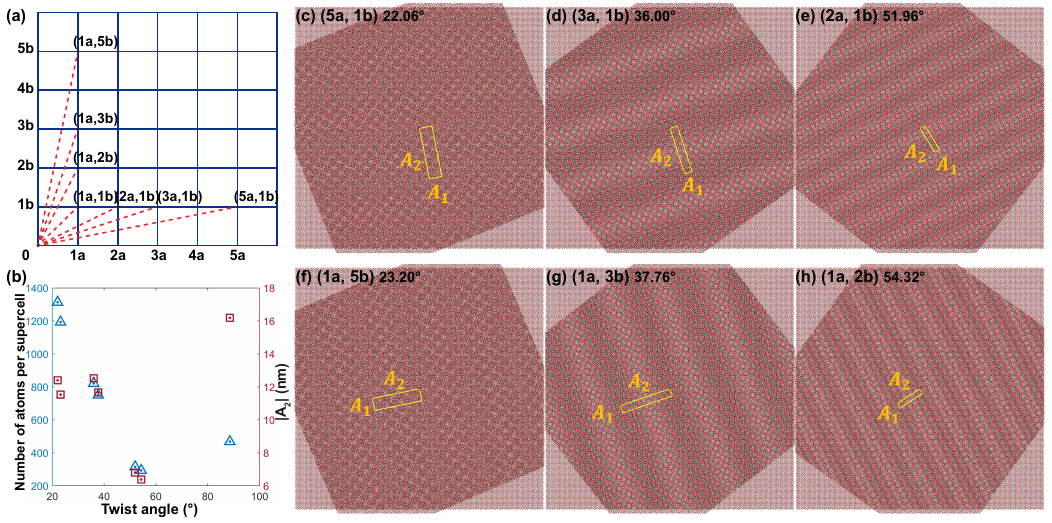}
\caption{ Critical angles generating 1D moir\'e patterns in twisted bilayer PdSe$_2$. (a) Schematic of the monolayer PdSe$_2$ lattice, showing seven lattice vectors (red dashed lines) connecting the origin $(0, 0)$ to lattice points $(ma, nb)$, where $m$ and $n$ are integers. These vectors define seven critical twist angles $i=1,\dots 7$ according to $\theta_\text{ci}=\mathrm{min}[2\arctan\left({ma}/{nb}\right),2\arctan\left({nb}/{ma}\right)]$. (b) The seven critical angles, along with their corresponding number of atoms per supercell and the 1D moir\'e wavelength $|A_2|$ between adjacent 1D channels. (c-h) 1D moir\'e patterns corresponding to the critical angles $\theta_\text{ci=2\dots 7}$ identified in (a).}
\label{fig:figure2}
\end{figure*}

The moir\'e superlattice constructed in this manner represents a highly special configuration among all possible twisted bilayer arrangements, and the system exhibits a behavior that depends very sensitively on the twist angle. As the twist angle approaches this critical value from either direction, the size of the moir\'e supercell and the number of atoms it contains diverge, as shown in Fig.~\ref{fig:figure1}(g). Only at the exact critical angle does the two-dimensional moir\'e pattern collapse into a one-dimensional pattern, accompanied by an abrupt drop in the number of atoms per supercell (see Figs. S4 and S5 in the SM for real-space moir\'e patterns and atom counts near the critical angles). This behaviour is significantly different from that of twist triangular or hexagonal lattices, for which the moir\'e cell size scales inversely with the twist angle. The total energy of the system at this critical configuration is significantly lower than that of nearby noncritical twist angles by about 20 meV per atom, suggesting that this state may be energetically favorable and can be potentially realized in experiments. Unfortunately, due to the rapid increase in system size near the critical angle, energy calculations in this regime become computationally intractable. 

The first critical angle $\theta_{\text{c1}}=88.53^\circ$ corresponds to the solution $\left(m, n, p, q\right) = \left(1, 1, 1, -1\right)$ in Eq.~\eqref{eq:sol1}. This approach can be generalized to identify additional critical twist angles and associated one-dimensional (1D) moir\'e configurations by considering lattice vectors of the form $(ma, nb)$ with \( m, n \in \{1, 2, 3, 4, 5\} \). As illustrated in Fig.~\ref{fig:figure2}(a), we highlight seven representative $(ma,nb)$ combinations and their corresponding supercell lattice vectors, which can be used to construct twisted superlattices exhibiting 1D moir\'e patterns. The real-space moir\'e structures and the associated critical twist angles for all these cases except the first critical angle are presented in Figs.~2(c–h). Moreover, the number of atoms per moir\'e cell diverges as the twist angle approaches each critical value (see Fig.~S4 in the SM). Interestingly, both the number of atoms per moir\'e cell and the corresponding 1D moir\'e wavelength $|A_2|$ at these critical angles show no clear twist angle dependence (Fig.~\ref{fig:figure2}(e)). This apparent lack of angle dependence arises because, in reciprocal space, the moir\'e reciprocal lattice vectors associated with different critical twist angles may originate from differences between the reciprocal lattices of the top and bottom layers beyond the first Brillouin zone, as illustrated in Fig.~S7 in the SM. The emergence of critical angles  and the formation of 1D moir\'e patterns is a general phenomenon for twisted rectangular lattices and can thus be extended to other 2D materials with rectangular symmetry, such as black phosphorus, WTe$_2$, GeSe, SnS, and related compounds \cite{PNAS_Ling2015BP,PRB2015BP}.

\begin{figure}
\includegraphics[width=1.0\linewidth,trim=0 0 0 0,clip]{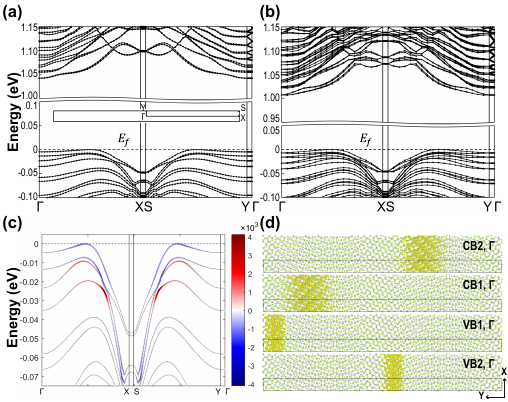}

\caption{DFT analysis of 1D moir\'e states for twisted bilayer PdSe$_2$ at $\theta_{\text{c1}}=88.53^o$. (a) Band structure without SOC along the high-symmetry path $\Gamma$-X-S-Y-$\Gamma$, ($E_f=0 eV$, dashed line). Inset: schematic of the first Brillouin zone. (b) Band structure with SOC. (c) Berry curvature distribution of the valence bands, the color scale on the bands and the color bar indicate the magnitude of the Berry curvature. (d) Real-space partial charge density distributions at the $\Gamma$ point for four bands near the fermi level (VB2,VB1,CB1,CB2), exhibiting 1D charge density wires localized along the 1D channels direction.}
\label{fig:figure3}
\end{figure}

\textit{Electronic properties}---To investigate the effects on the electronic properties by the 1D moir\'e structure, we calculate the DFT band structures for twisted bilayer PdSe$_2$ at $\theta_{c1}=88.53^\circ$ and the results are shown in Fig.~\ref{fig:figure3}.
The system exhibits an indirect band gap of 1.04 eV (Fig.~\ref{fig:figure3}(a)). Near the Fermi level, the valence and conduction bands show pronounced dispersion along the $k_x$ direction, while demonstrating a 1D flat-band behavior (i.e., absence of dispersion) along the $k_y$ direction (further details in Fig.~S9) \cite{NC_Kennes_2020_1DFlatInGeSe,Li2021AnisoFlatBand,yu2023evidence,Khoury2024Toward1D,gao20251Dflat}.
At higher energies in the conduction bands, dispersion reappears along the 
$k_y$ direction, whereas the valence band retains its 1D flat-band character over a wider energy range, indicating that it is more strongly modulated by the moir\'e potential.
Notably, the two highest valence bands (VB1 and VB2) possess an extremely narrow energy bandwidth ($<$ 50 meV), and exhibit prominent flat-band features near the $\Gamma$ point of the Brillouin zone. These characteristics suggest that this system could exhibit  strong correlation effects \cite{naik2018ultraflatbands,NC_Kennes_2020_1DFlatInGeSe,devakul2021magic,vitale2021flat,li2024TuningFlat}. In contrast, near the X point, the moir\'e potential induces significant band gap openings.\\
\indent The pronounced flatness of the upper valence bands originates from their highly localized charge density along the 1D moir\'e channel direction (see Fig.~\ref{fig:figure3}(d)).
When spin-orbit coupling is included, the band structure remains degenerate at the $\Gamma$ and X points, as well as along the XS and $\Gamma$Y segments (i.e., along the $k_y$ direction), while significant band splitting occurs everywhere else (Fig.~\ref{fig:figure3}(b)). This momentum-dependent splitting indicates the existence of substantial Rashba spin-orbit coupling (RSOC) in the system\cite{NM2015NewRSOC,NRP2022rashba}. In such a 1D moir\'e structure, the coexistence of exceptionally flat bands and strong RSOC provides a promising platform for correlated electron physics, where the conventional Fermi liquid description may break down due to lowered dimensionally, and gives way to Luttinger liquid behaviour \cite{Wen1990MetallicNFermiLuttinger,2001two2DnFermiLuttinger,Na_Wang_2022_1DLLs}. Combined with the strong RSOC of the system we establish that twisted bilayer PdSe$_2$ at $\theta_{c1}=88.53^\circ$ provides a unique inroad into the interplay of Luttinger liquid and RSOC physics with a particularly promising direction being the stabilization of  Majorana zero-energy modes at the channels' ends when proximitized by a conventional supercondcutor \cite{Fu2008MajoranaFermions,Oreg2010MajoranaBoundStates,Lutchyn2010MajoranaFermions,chen2024chirality,tanaka2024MZMtheory}.\\
\indent Fig.~\ref{fig:figure3}(c) reveals a nonzero Berry curvature distribution in the top valence bands, indicating significant interband topological obstruction. Notably, at the band crossing point, the second valence band exhibits a positive peak in Berry curvature, while the third valence band exhibits a negative peak. This indicates the presence of topologically nontrivial structures in the bands \cite{PRL2015BerryCurvatureDi,NP2022BCDinMoire}. A positive Berry curvature peak also appears in the fourth valence band, comparable in magnitude but opposite in sign to that of the third band, suggesting strong interband topological connection between them. In this one-dimensional (1D) moir\'e system, these topological features imply that the system may host symmetry-protected edge states or enable topological charge pumping under adiabatic modulation \cite{PRX2017BCNonlocalTranport,NP2017ExperiBC}. To study this in detail any low energy description requires a faithful representation of the topologically obstructed bands, requiring at least a two-band description, which is left as a subject of future study.\\
\textit{Note added}.---Recently, we became aware of a related study on moir\'e collapse and Luttinger liquids in twisted anisotropic homobilayers \cite{desousa2025moirecollapseluttingerliquids}, which is complementary to our investigation of topological 1D flat bands and critical angles in twisted rectangular lattices.

\textit{Acknowledgments}---LX acknowledges supported by the National Key Research and Development Program of China (Grant No. 2022YFA1403501), Guangdong Basic and Applied Basic Research Foundation (Grant No. 2022B1515120020), the National Natural Science Foundation of China (Grant No. 62341404), Hangzhou Tsientang Education Foundation and the Max Planck Partner group programme.
LW acknowledges support from the National Key Projects for Research and Development of China (Grant Nos. 2021YFA1400400, 2022YFA1204700) and Natural Science Foundation of Jiangsu Province (Grant Nos. BK20220066 and BK20233001).
DMK acknowledges funding by the DFG  within the Priority Program SPP 2244 ``2DMP'' -- 443274199.
AR acknowledges support by the European Research Council (ERC-2015-AdG694097), ERC Synergy Grant Agreement No. 101167294 (UnMySt), the Cluster of Excellence ‘Advanced Imaging of Matter' (AIM), Grupos Consolidados (IT1249-19) and Deutsche Forschungsgemeinschaft (DFG) -- SFB-925 -- project 170620586. The Flatiron Institute is a division of the Simons Foundation. 
DMK and AR acknowledge support by the Max Planck-New York City Center for Nonequilibrium Quantum Phenomena.

\nocite{*}

\bibliography{main_v1}

\end{document}